\documentclass[prd,preprint,aps,dvips]{revtex4}
\usepackage[english]{babel}
\usepackage{amscd}
\usepackage{epsfig}
\usepackage{tabularx}
\usepackage{graphicx}
\usepackage{latexsym}
\usepackage{amsmath}
\usepackage{amsfonts}
\usepackage{amssymb}
\usepackage[latin1]{inputenc}
\usepackage{times}
\usepackage[T1]{fontenc}
\usepackage{latexsym}
\usepackage{graphics}
\usepackage{verbatim}
\usepackage[absolute]{textpos}
\usepackage{wrapfig,times}
\usepackage{amsthm}
\usepackage{setspace}

\newcommand{\be}{\begin{equation}}
\newcommand{\ee}{ \end{equation}}
\newcommand{\ben}{\begin{eqnarray}}
\newcommand{\een}{\end{eqnarray}}

\begin{document}

\title{Thermal and electrical properties of a solid through Fibonacci oscillators}

\author{André A. Marinho$^{a}$, Francisco A. Brito$^{a}$, Carlos Chesman$^{b}$}

\affiliation{$^{a}$ Departamento de Física, Universidade Federal de
Campina Grande, 58109-970 Campina Grande, Paraiba, Brazil
\\
$^{b}$ Departamento de Física Teórica e Experimental, Universidade Federal do Rio 
Grande do Norte, 59078-970 Natal, RN, Brazil. 
}

\begin{abstract} 
We investigate the thermodynamics of a crystalline solid  applying $q$-deformed algebra of Fibonacci oscillators through the generalized Fibonacci sequence of two real and independent deformation	parameters $q_1$ and $q_2$. We based part of our study on both Einstein and Debye models, exploring primarily $(q_1,q_2)$-deformed thermal and electric conductivities as a function of Debye specific heat. The results revealed that $q$-deformation acts as a factor of disorder or impurity, modifying the characteristics of a crystalline structure. Specially, one may find the possibility of adjusting the Fibonacci oscillators to describe the change of thermal and electrical conductivities of a given element as one inserts impurities. Each parameter can be associated to different types of deformations such as disorders and impurities. 
\end{abstract}


\maketitle


\section{Introduction}

The interaction between atoms allows propagation of elastic waves in the solid medium which can be both transverse and
longitudinal. If the oscillations of the atoms around the equilibrium positions are small, which should occur at lower temperatures, 
the potential energy of interaction can be approximated by a quadratic form of the displacements of atoms from their 
equilibrium positions. A crystalline solid, whose atoms interact according to this potential, is called a harmonic solid. 
In harmonic solids, elastic waves are harmonics and the normal modes of vibration in crystalline solids \cite{kit2}. 
A large number of phenomena involve quantum mechanical motion, in particular thermally-activated particles, obeying the
$T^{3}$ law. Thermal excitations in the system are responsible for phonon excitation \cite{patt,hua}. 

The study conducted by Anderson, Lee and Elliot \cite{ande, lee, ell} shows that the presence of defects or impurities in a 
crystal modifies the electrostatic potential in their neighborhood, breaking the translational symmetry of the periodic potential. 
This perturbation can produce electronic wave functions located near the impurity, ceasing to be propagated throughout the crystal. 

A possible way to generate a deformed version of the classical statistical mechanics consists in replacing the Gibbs-Boltzmann
 distribution by a deformed version. In this respect it is postulated a form of deformed entropy \cite{tsalis} which implies a generalized theory of thermodynamics.

We apply the $q$-deformation in the models of Einstein and Debye \cite{bri2,bri3,bri6}, and our results show that the factor 
$q$ acts as an impurity, modifying the thermodynamic quantities such as entropy, specific heat, 
thermal conductivity, etc. 

In this work we insert the given parameters of deformation $q_1$ and $q_2$, called Fibonacci oscillators \cite{arik2}, 
which is a formalism recently proposed in the $q$-calculation that has been investigated in \cite{aba, amg1, bri4}. They provide a unification of quantum oscillators with quantum groups, keeping the degeneration property of the spectrum invariant under the symmetries of the quantum group \cite{bie1, mac, fuc, erz, anat}. The quantum algebra with two deformation parameters may have a greater flexibility when it comes to applications in realistic phenomenological physical models  \cite{dao,gong} and may increase interest in physical applications.

The paper is organized as follows. In Sec.~\ref{alg} we present the deformed algebra. In Sec.~\ref{aof}
we apply the Fibonacci oscillators, in the Einstein \ref{sed} and Debye \ref{sdd} models, and finally,
in Sec.~(\ref{con}) we make our final comments. 

\section{Algebra of the Fibonacci oscillators}
\label{alg}

It is well-known that the generalization of integers in general is given by a sequence. A basic procedure in $q$-algebra \cite{jac1} 
is a generalization of integers. Two well-known ways to describe a sequence are the arithmetic and geometric progressions. 
A simple generalization that encompasses both of is the Fibonacci sequence, which as we know is a linear combination where the 
third number is the sum of two predecessors, and so on.  Here, the numbers are in that sequence of generalized Fibonacci oscillators, 
where new parameters $(q_1,q_2)$ are introduced. Thus, the generalized spectrum may be given by the whole Fibonacci number. 

The algebraic symmetry of the quantum oscillator is defined by the Heisenberg algebra in terms of the annihilation and creation 
operators $c$, $c^{\dagger}$, respectively, and the number operator $N$ \cite{aba,lav1} via
\be c_i c_{i}^\dagger - q_1^{2}c_{i}^\dagger c_i = q_2^{2n_i}\qquad\mbox{and}\qquad c_i c_{i}^\dagger - 
q_2^{2}c_{i}^\dagger c_i = q_1^{2n_i},\ee
\begin{equation}[N,c^{\dagger}] = c^{\dagger}, \qquad\qquad [N,c] = -c.\end {equation}
In addition, the operators obey the relations 
\begin{eqnarray}c^\dagger c=[N],\;\;\qquad cc^{\dagger} = [1+N], \end {eqnarray}
\be [1+n_{i,q_1,q_2}] = q_1^{2}[n_{i,q_1,q_2}]+q_2^{2n_i},\;\quad\mbox{or}\quad\; [1+n_{i,q_1,q_2}] = q_2^{2}[n_{i,q_1,q_2}]
+q_1^{2n_i}.\ee
The oscillator \cite{bie1, mac} allows us to write the $(q_1,q_2)$-deformed Hamiltonian \cite{anat} as follows
\begin{equation} \label{e1}{\cal H} = \frac{1}{2}\Big{\{c,c^{\dagger}\Big\}}.\end{equation}
The Fibonacci \textit{basic number} is defined by \cite{arik}
\be \label{e49}[n_{i,q_1,q_2}] = c_{i}^\dagger c_{i} = \frac{q_2^{2n_i}-q_1^{2n_i}}{q_2^2-q_1^{2}},\ee
where $q_1$ and $q_2$ are real positive and independent parameters of deformation.

\section{Application of Fibonacci Oscillators}
\label{aof}
\subsection{($q_1,q_2$)-deformed Einstein solid}
\label{sed}

We consider the solid in contact with a thermal reservoir at temperature $T$, where we have $n_j$ labeling the $j$-th oscillator
Given a microscopic state $\{n_j\}=\{n_1,n_2,\ldots,n_N\}$, the energy of this state can be written as,
\begin{equation}\label{eq1} E\{n_j\} = \displaystyle\sum_{j=1}^{\infty}\left(n_j+\frac{1}{2}\right)\hbar\omega_E,\end{equation}
where $\omega_E$ is the Einstein frequency characteristic. 
We can obtain $(q_1,q_2)$-deformed energies from the definition of the Hamiltonian (\ref{e1}), and the definitions provided earlier, 
\be \label{eq2}E_{n_{i,q_1,q_2}} = \frac{\hbar\omega_E}{2}\Big([n_{i,q_1,q_2}]+[n_{i,q_1,q_2}+1]\Big)=
\frac{\hbar\omega_E}{2}+\frac{\hbar\omega_E\left(2\ln(q_2)-2\ln(q_1)\right)n}{q_2^2-q_1^2},\ee
and when $q_1=q_2=1$, we recover the usual spectrum
\be E_n = \frac{\hbar\omega_E}{2}\left(2n+1\right).\ee
With the result of the Eq.(\ref{eq2}), we can rewrite the partition function in the form, 
\begin{equation} \Xi_{(q_1,q_2)} = \Bigg\{\displaystyle\sum_{n=0}^{\infty}\exp\left[-\beta E_{n_{i,q_1,q_2}} 
\right]\Bigg\}^N = \Xi_{(1,q_1,q_2)}^{N}, \end{equation}
where 
\be \label{eq3}\Xi_{(1,q_1,q_2)} = \frac{q_{1}^{2\alpha}}{\exp\left(\frac{\alpha(q_{1}^2-q_{2}^2)}{2}\right)(q_{1}^{2\alpha}-
q_{2}^{2\alpha})}.\ee
We define a $(q_1,q_2)$-deformed Einstein function $E(\alpha)_{q_1,q_2}$, 
\ben \label{eq4}E(\alpha)_{q_1,q_2}=\left(\frac{2\alpha q_1q_2[\ln(q_1)-\ln(q_2)]}{(q_{1}^{2\alpha}-q_{2}^{2\alpha})}\right)^{2},
\qquad\mbox{where}\qquad \alpha=\left(\frac{\Theta_E}{T(q_{1}^2-q_{2}^2)}\right).\een
As one knows $\Theta_E$ is the Einstein temperature, defined by 
\be \Theta_E=\left(\frac{\hbar\omega_E}{\kappa_B}\right),\ee
and $\kappa_B$ is the Boltzmann constant. When $q_1=q_2\to 1$, we have the undeformed  function 
\ben E(\alpha^*)=\frac{\exp(\alpha^*)\alpha^{2*}}{(\exp{\alpha^*}-1)^2}, \qquad\;\mbox{where}\qquad\alpha^*=\frac{\Theta_E}{T}.\een

We can determine ($q_1,q_2$)-deformed the Helmholtz free energy per oscillator and entropy, respectively 
\ben \label{eq5}f_{q_1,q_2} = -\frac{1}{\beta}\;\displaystyle\lim_{N\to\infty}\frac{1}{N}\ln\Xi_{(q_1,q_2)} = -\kappa_{B} T
\ln\left[\frac{q_{1}^{2\alpha}}{\exp\left(\frac{\alpha(q_{1}^2-q_{2}^2)}{2}\right)(q_{1}^{2\alpha}-q_{2}^{2\alpha})}\right].\een

\ben S_{q_1,q_2} &=& -\frac{\partial f_{q_1,q_2}}{\partial T} = \kappa_{B}\Biggl\{\ln\Bigg[\frac{q_1^{2\alpha}}
{\exp\Big(\frac{\alpha(q_1^2-q_2^2)}{2}\Big)(q_1^{2\alpha}-q_2^{2\alpha})}\Bigg]+\nonumber\\
&+&\Bigg[\frac{2\alpha q_2^{2\alpha}(\ln(q_1)-\ln(q_2))}{(q_1^{2\alpha}-q_2^{2\alpha})}
+\frac{\alpha(q_1^2-q_2^2)}{2}\Bigg]\Biggl\}.\een

\begin{figure}[htb]
\centerline{
\includegraphics[{angle=90,height=7.0cm,angle=270,width=7.0cm}]{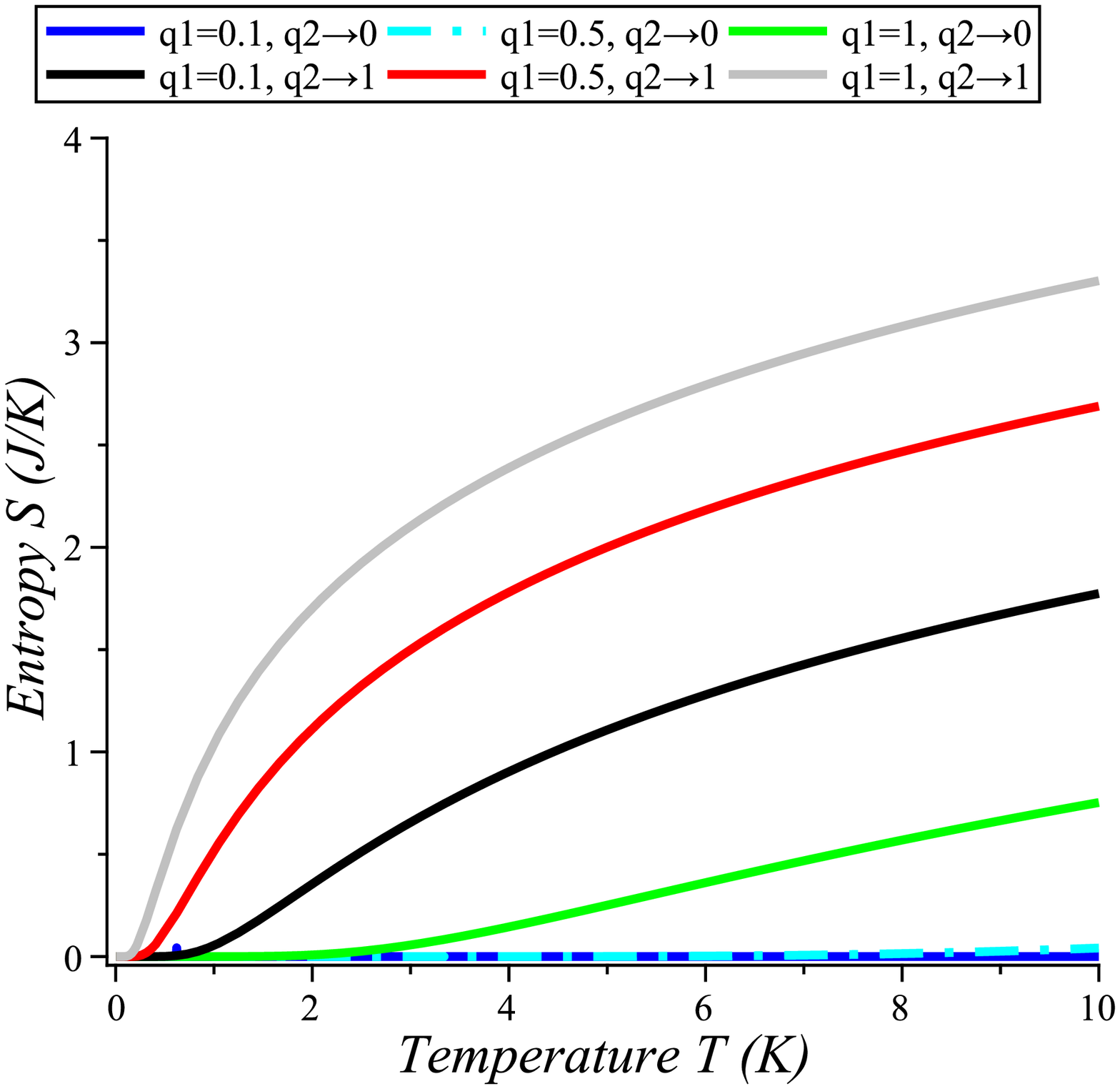}
\includegraphics[{angle=90,height=7.0cm,angle=270,width=7.0cm}]{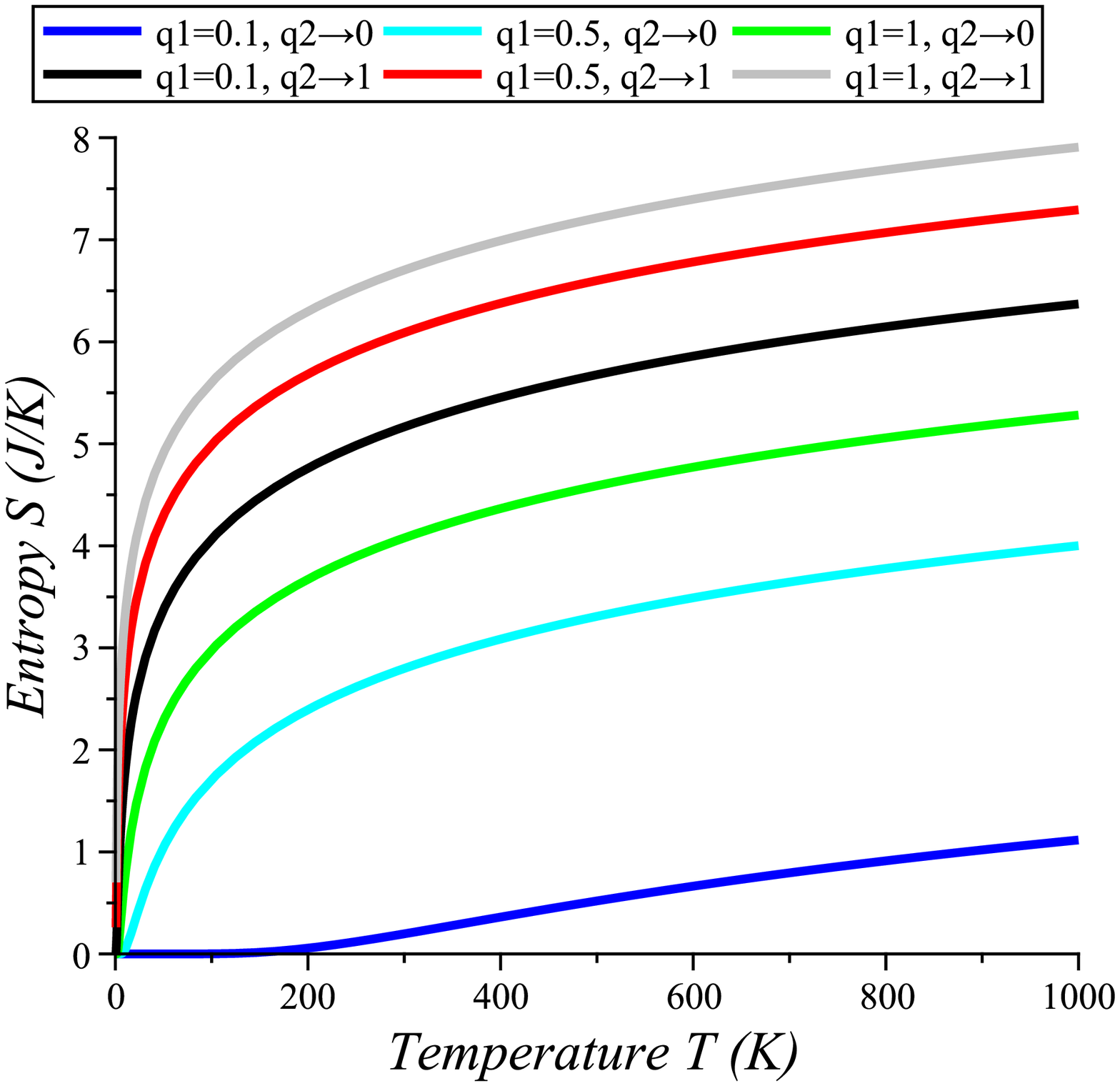}
}\caption{\small{${q_1,q_2}$-deformed entropy $S_{q_1,q_2}$ vs temperature T in the following intervals: 
(T=$0,\cdots, 10 K$) \textbf{(left)} and (T=$0,\cdots, 1000 K$) \textbf{(right)}}}\label{gráficos 14}
\end{figure}

In Fig.~(\ref{gráficos 14}) is shown the behavior of the entropy as a function of temperature variation. We observe that all the 
curves have the same behavior at low temperature. However, as the temperature increases  the role of the $q$-deformation becomes much
more evident. For instance, note that $q_2$ tends do decrease the entropy more than $q_1$. As we anticipated, these parameters can play different roles.
While one can affect disorders the other may control impurities. 

Now we determine ($q_1,q_2$)-deformed specific heat, and we can do it by inserting the Einstein function $E(\alpha)_{q_1,q_2}$, 
defined by Eq.(\ref{eq4}), into equation below
\begin{equation} c_{V_{q_1,q_2}}(T) = T\left(\frac{\partial S_{q_1,q_2}}{\partial T}\right) = \kappa_{B}\,
\left(\frac{2\alpha q_1q_2[\ln(q_1)-\ln(q_2)]}{(q_1^{2\alpha}-q_2^{2\alpha})}\right)^2.\end{equation}
\begin{equation} \label{eq6} c_{V_{q_1,q_2}}(T) = 3\kappa_{B}E(\alpha)_{q_1,q_2}.\end{equation}
\begin{figure}[htb]
\centerline{
\includegraphics[{angle=90,height=9.0cm,angle=270,width=5.2cm}]{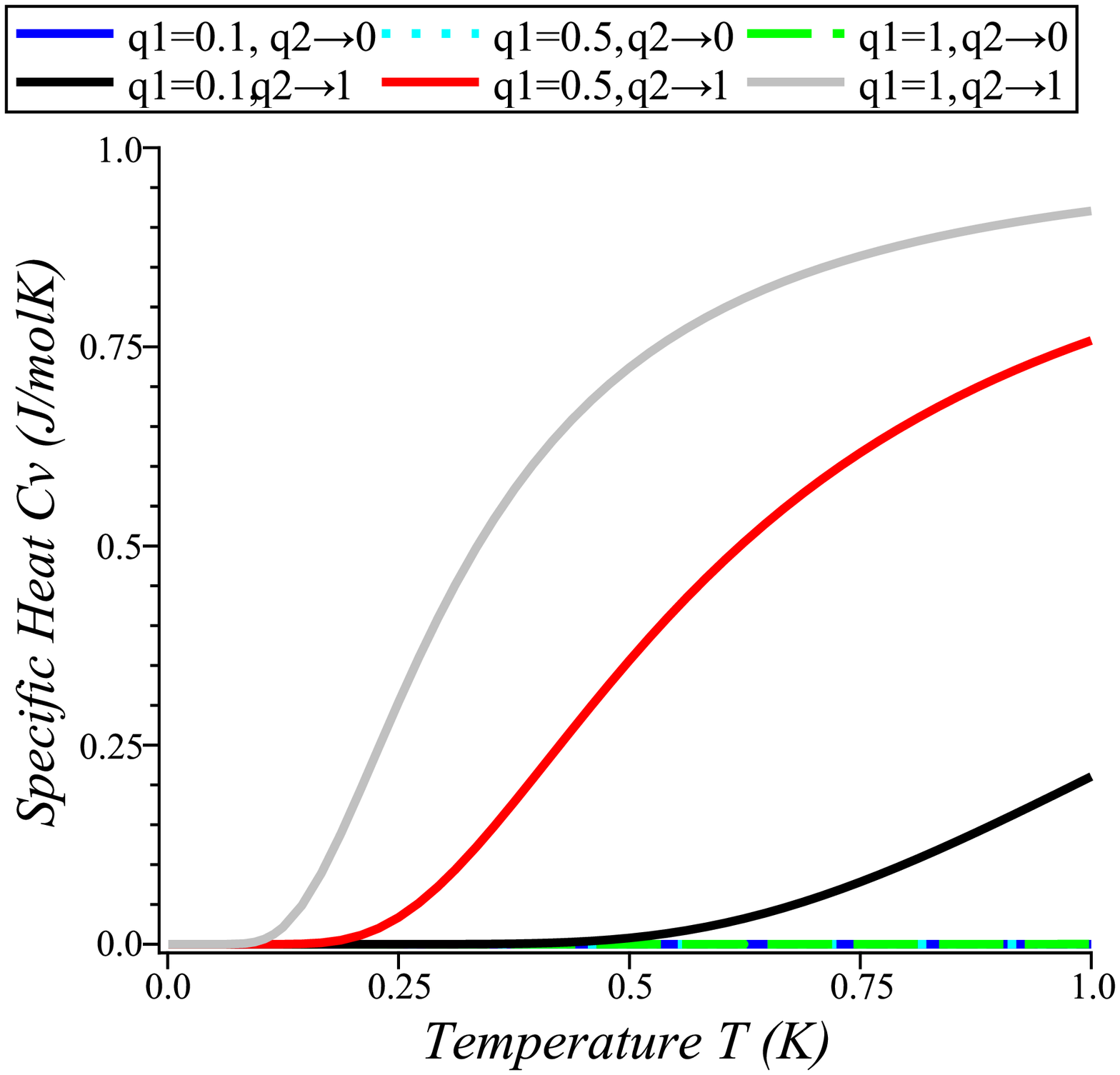}
\includegraphics[{angle=90,height=9.0cm,angle=270,width=5.2cm}]{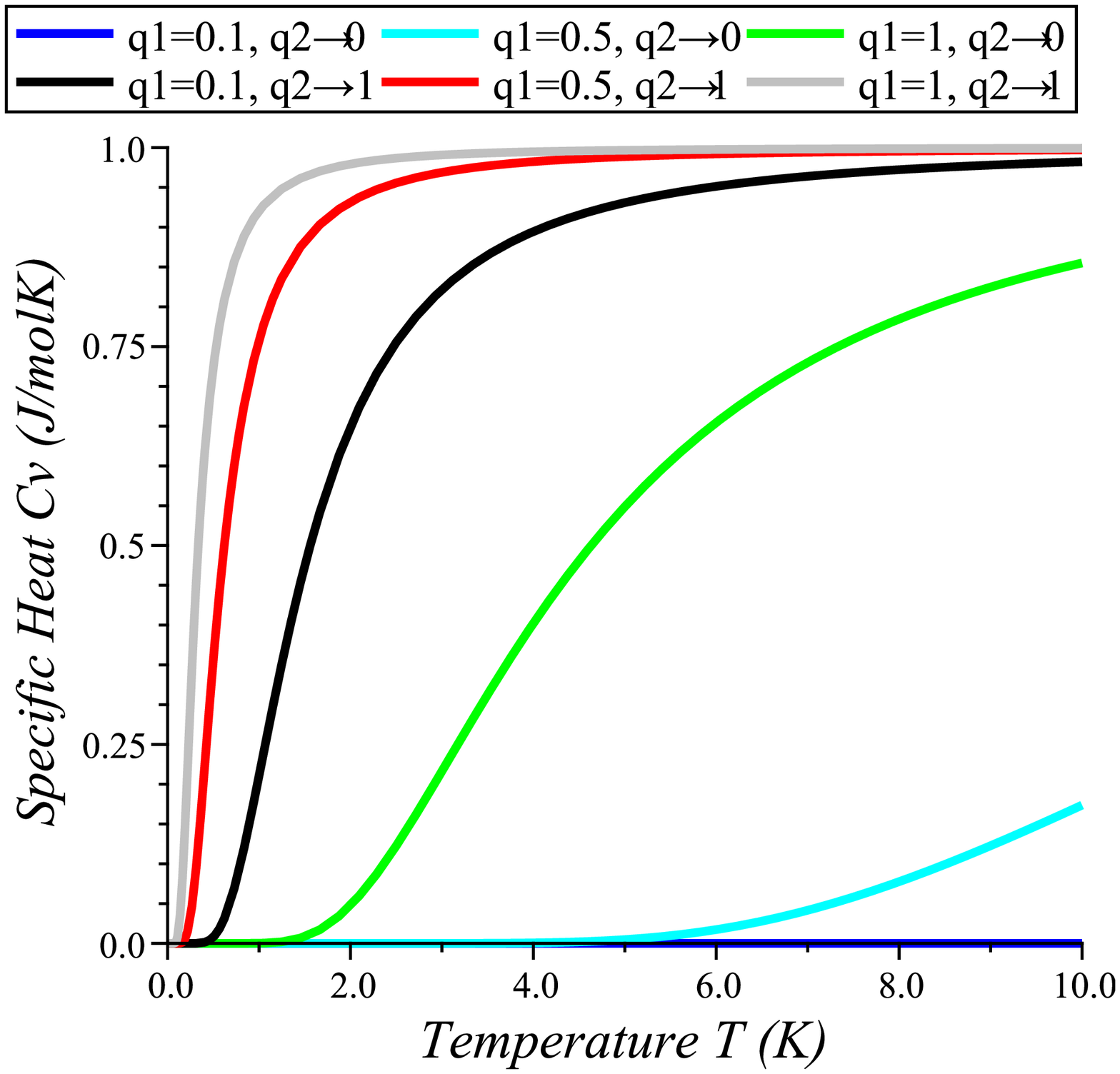}
\includegraphics[{angle=90,height=9.0cm,angle=270,width=5.2cm}]{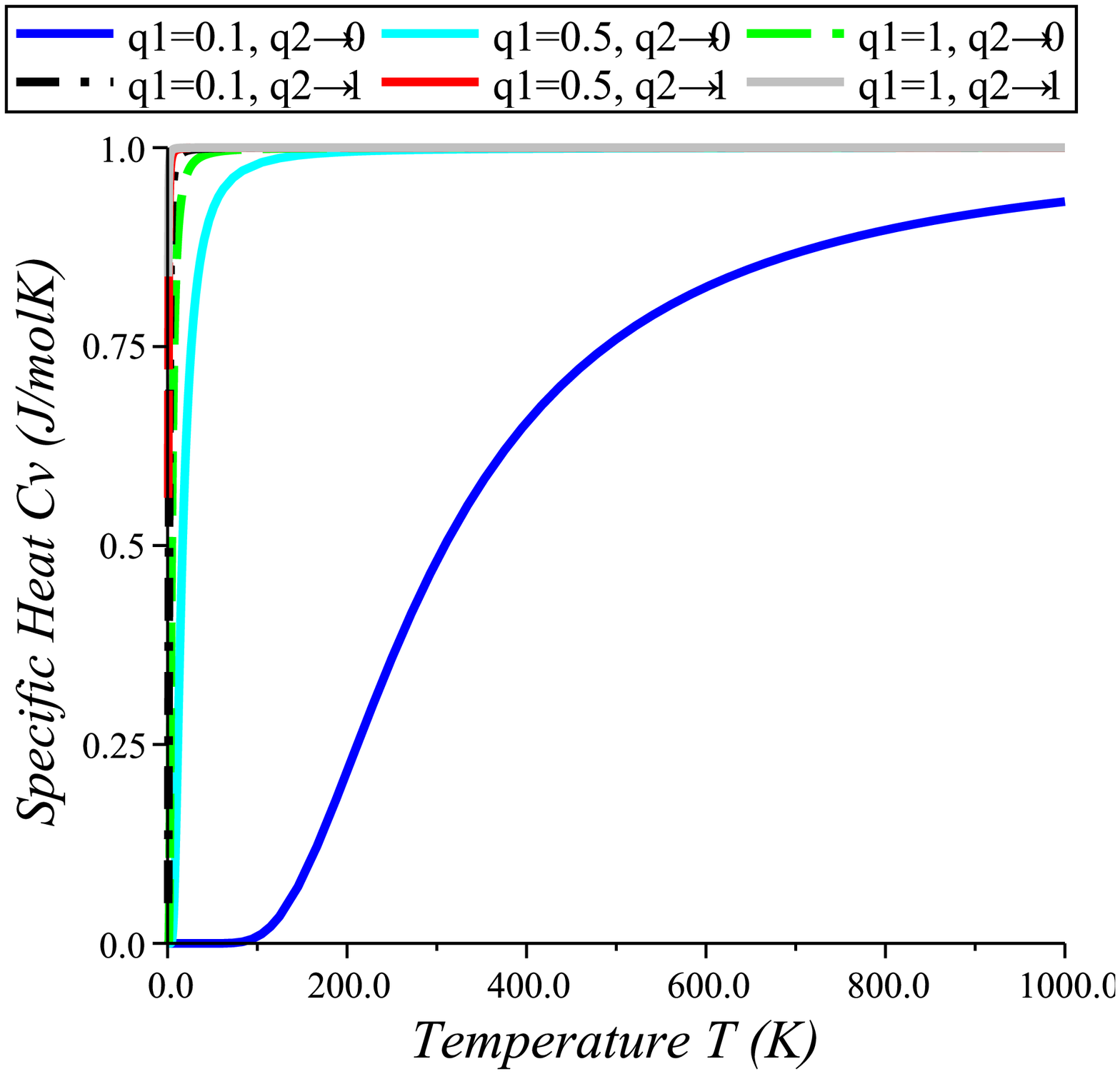}
}\caption{\small{${q_1,q_2}$-deformed specific heat $c_{V_{q_1,q_2}}$ vs temperature T in the following intervals: 
(T=$0,\cdots, 1 K$) \textbf{(left)}, (T=$0,\cdots, 10 K$) \textbf{(center)} and (T=$0,\cdots, 1000 K$) 
\textbf{(right)}.}}\label{gráficos 15}
\end{figure}
The complete behavior is depicted in Fig.~(\ref{gráficos 15}). One should note that when $T\gg\Theta_E$, and thus the ratio 
$\alpha=\frac{\Theta_E}{T}\ll 1$, and $\Theta_E$ around $100 K$ for common crystals, one recovers the classical result 
$c_{V_{q_1,q_2}}\to 3\kappa_B$, known as the Dulong-Petit law. 
However, for sufficiently low temperatures, where $T\ll\Theta_E$ and therefore $\alpha\gg 1$, specific heat decreases exponentially with 
temperature \cite{patt}, as 
\begin{equation}\label{ex9} c_V \rightarrow \kappa_B 
\left(\frac{\theta_E}{T}\right)^2\exp\left(-\frac{\theta_E}{T}\right). \end{equation} 

In general, the invariance of specific heat at high temperatures and its decrease at low temperatures show that the 
Einstein model is in agreement with experimental results. However, at sufficiently low temperatures, specific heat does 
not experimentally follow the exponential function given in Eq.(\ref{ex9}). 
As for the $(q_1,q_2)$-deformed case we see a significant change in the curves  at intermediate temperatures.

\subsection{($q_1,q_2$)-deformed Debye solid}
\label{sdd}

Corrections of Einstein model are given by the Debye model, allowing us to integrate from a continuous spectrum of
frequencies up to the Debye frequency $\omega_D$, giving the total number of normal modes of vibration \cite{patt,hua,kit2}
\begin{equation} \displaystyle\int_{0}^{\omega_D}g(\omega)d\omega = 3N, \end{equation}
where $g(\omega)d\omega$ denotes the number of normal modes of vibration whose frequency is in the range $(\omega, \omega+d\omega)$. 
The function $g(\omega)$, can be given in terms of the Rayleigh expression as follows 
\begin{equation} \label{eq9}8\pi\left(\frac{1}{\lambda}\right)^{2}d\left(\frac{1}{\lambda}\right) = \frac{\omega^2 d\omega}
{\pi^{2}c^{3}},\end{equation}
where $c$ is the speed of light and $\lambda$ wavelength. The expected energy value of the Planck oscillator with frequency 
$\omega_s$ is 
\ben \label{eq10}
\langle E_s\rangle = \frac{\hbar\omega_s}{\exp\left(\frac{\hbar\omega_s}{\kappa_{B}T}\right)-1}.\een
Using Eqs.(\ref{eq9}) and (\ref{eq10}), we obtain the energy density associated with the frequency range $(\omega,\omega+d\omega)$, 
\begin{equation} u(\omega)d\omega = \frac{\hbar}{\pi^2 c^3}\frac{\omega^3d\omega}{\exp\left(\frac{\hbar\omega}
{\kappa_B T} \right)-1}.\end{equation}
To obtain the number of photons between $\omega$ and $\omega+d\omega$, one makes use of the volume of the region on the phase space 
\cite{patt}, which results in 
\begin{eqnarray} \label{eq11} g(\omega)d\omega \approx \frac{2V}{h^3}\left[4\pi\left(\frac{\hbar\omega}{c}\right)^2 
\left(\frac{\hbar d\omega}{c}\right)\right] = \frac{V\omega^{2}d\omega}{\pi^{2}c^{3}}. \end{eqnarray}
Thus, replacing Eq.(\ref{eq11}) into Eq.(\ref{eq9}), we can write the specific heat for any temperature. 
We now apply ($q_1,q_2$)-deformation in the same way as in Eq.(\ref{eq6}),
\begin{equation}  c_{V_{q_1,q_2}}(T) = 3\kappa_{B}D(\alpha_{0_{q_1,q_2}}), \end{equation}
where $D(\alpha_{0_{q_1,q_2}})$ is the ($q_1,q_2$)-deformed Debye function, defined by 
\ben \label{eq12} D(\alpha_{0_{q_1,q_2}}) = \frac{3}{(\alpha_{0_{q_1,q_2}})^3} \int_{0}^{\alpha_{0_{q_1,q_2}}} 
\frac{\alpha^4\exp{(\alpha)}}{[\exp(\alpha)-1]^2}d\alpha,\qquad\;{\alpha_{0_{q_1,q_2}}} = \frac{\hbar\omega_{D_{q_1,q_2}}}
{\kappa_{B}T} = \frac{\theta_{D_{q_1,q_2}}}{T},\een
\be \omega_{D_{q_1,q_2}} = 2\omega_D\left(\frac{\ln(q_2)-\ln(q_1)}{q_2^2-q_1^2}\right), \ee 
where $\omega_{D_{q_1,q_2}}$ and $\theta_{D_{q_1,q_2}}$, are the $(q_1, q_2)$-deformed Debye frequency and temperature, 
respectively, and $\omega_D$ is the Debye frequency characteristic. Integrating Eq.(\ref{eq12}) by parts one finds 
\be D(\alpha_{0_{q_1,q_2}}) = -\frac{3{\alpha_{0_{q_1,q_2}}}}{\exp({\alpha_{0_{q_1,q_2}}})-1} + 
\frac{12}{\alpha_{0_{q_1,q_2}}^3}\int_{0}^{\alpha_{0_{q_1,q_2}}} {\frac{\alpha^3 d\alpha}{\exp(\alpha)-1}},\ee
which can be integrated out to give the full expression 
\ben \label{eq12.5} D(\alpha_{0_{q_1,q_2}})&=&-\frac{3{\alpha_{0_{q_1,q_2}}}}{\exp({\alpha_{0_{q_1,q_2}}})-1}+\frac{12}
{\alpha_{0_{q_1,q_2}}^3}\Bigg\{-\frac{\pi^4}{15}-\frac{\alpha_{0_{q_1,q_2}}^4}{4}+\alpha_{0_{q_1,q_2}}^3
\ln[1-\exp(\alpha_{0_{q_1,q_2}})]\nonumber\\
&+&3\alpha_{0_{q_1,q_2}}^2{\rm Li}_2[\exp(\alpha_{0_{q_1,q_2}})]-6\alpha_{0_{q_1,q_2}}{\rm Li}_3[\exp(\alpha_{0_{q_1,q_2}})]+6{\rm Li}_4
[\exp(\alpha_{0_{q_1,q_2}})]\Bigg\},\een
where 
\be {\rm Li}_n(z)=\displaystyle\sum_{k=0}^{\infty}{\frac{z^n}{k^{n}}},\ee 
is the polylogarithm function. For $T\gg\theta_{D_{q}}$,  $\alpha_{0{(q_1,q_2)}}\ll 1$, then the function $D(\alpha_0)_{q}$ 
can be expressed in a power series in $\alpha_{0_{q}}$ 
\begin{equation} D(\alpha_{0})_{q} = 1 - \frac{\alpha_{0_{q}}^2}{20} + \cdots \end{equation}
so that for 
\begin{eqnarray} T\to\infty\;,\qquad c_{V{q}}\to 3\kappa_B.
\end{eqnarray}
On the other hand, for $T\ll \theta_{D_{q}}$, $\alpha_{0_q}\gg 1$, then we can write function $D(\alpha_0)_q$ as 
\begin{equation} 
\frac{12}{\alpha_{0_q}^{3}}\displaystyle\int_{0}^{\infty}{\frac{\alpha^3d\alpha}{\exp(\alpha)-1}+O[\exp(-\alpha_{0_q})]}, 
\end{equation}
\begin{equation}\approx \frac{4\pi^4}{5\alpha_{0_q}^{3}} = \frac{4\pi^4}{5}\left(\frac{T}{\theta_{D_{q}}}\right)^3. 
\end{equation}

Thus, as in the usual Debye solid, the low-temperature specific heat in a q-deformed Debye solid is proportional to $T^3$, 
due to phonon excitation, a fact that is in agreement with experiments. 
Thus, let us express the ($q_1,q_2$)-deformed specific heat for low temperatures as follows: 
\begin{eqnarray} c_{V_{q_1,q_2}} = \frac{12\pi^4\kappa_B}{5}\left(\frac{T}{\theta_{D_{q_1,q_2}}}\right)^3 = 1944\left(\frac{T} 
{\theta_{D_{q_1,q_2}}}\right)^3 \frac{J}{mol K}. \end{eqnarray}

For the ($q_1,q_2$)-deformed case one can observe the changes that occur with Debye temperature, specific heat, thermal and 
electrical conductivies. By using the relationship established for thermal conductivity $(\kappa)$ --- see \cite {zim}, we obtain 
\begin{equation} \label{eq13}\kappa = \frac{1}{3}C_{V}v l, \end{equation} 
where $v$ is the average velocity of the particle, $C_{V}$ is the molar heat capacity and $l$ is the space between particles. 
We can deduce a relationship between the thermal and electrical $(\sigma)$ conductivities through the elimination of $l$ 
(as $\sigma=\frac{ne^2l}{mv}$, where $m$ is the electron mass, $n$ is the number of electrons per volume unit and $e$ 
is the electron charge), such that
\begin{equation} \frac{\kappa}{\sigma}=\frac{1}{3}\frac{C_{V}mv^2}{ne^2}.\end{equation}
In a classical gas the average energy of a particle is $\frac{1}{2}mv^2=\frac{3}{2}\kappa_{B}T$, 
whereas the heat capacity is $\frac{3}{2}n\kappa_{B}$, so that
\be\label{eq14} \frac{\kappa}{\sigma}=\frac{3}{2}\left(\frac{\kappa_B}{e}\right)^2T. \ee
The ratio $\frac{\kappa}{\sigma T}$ is called the \textit{Lorenz number} and should be a constant, independent of 
the temperature and the scattering mechanism. This is the famous Wiedemann-Franz law, which is often well 
satisfied experimentally, and the Lorenz number correctly given \cite{zim}. By using the $(q_1,q_2)$-deformed 
relations presented above, we start from Eqs.(\ref{eq13}) and (\ref{eq14}) to determine the 
important relations for $(q_1,q_2)$-deformed thermal and electrical conductivities
\begin{eqnarray}\kappa_{q_1,q_2}=\frac{\kappa c_{V_{q_1,q_2}}}{c_{V}} \qquad\qquad \mbox{and} \qquad\qquad 
\sigma_{q_1,q_2}=\frac{\kappa_{{q_1,q_2}}\sigma}{\kappa}.\end{eqnarray}

Recall that to compute these deformed quantities in terms of the specific heat $c_{V_{q_1,q_2}}$ we make use of Eq.(\ref{eq12.5}) 
and its suitable limits. We present in Tab.~(\ref{tab2}), changes that occur with the Debye temperature, specific heat, 
thermal conductivity and electrical conductivity, of some chemical elements. 

\begin{table}[hbt]
\centering
\scriptsize
\begin{tabular}{|c||c|c|c||c|c|c||c|c|c||c|c|c|c|}
\hline
{Element} & \multicolumn {3} {c||} {$\theta_{D_{q}}^{(a)}$} & 
\multicolumn {3} {c||} {$C_{V_q}^{(b)}$} & 
\multicolumn {3} {c||} {$\kappa_q ^{(c)}$} & 
\multicolumn {3} {c|} {$\sigma_q ^{(d)}$} \\
\cline{2-13} 
& {$q_{2}$=1} & {$q_{2}$=0.5} & {$q_{2}$=0.1} & {$q_{2}$=1} & {$q_{2}$=0.5} & {$q_{2}$=0.1} & {$q_{2}$=1} & {$q_{2}$=0.5} 
& {$q_{2}$=0.1} & {$q_{2}$=1} & {$q_{2}$=0.5} & {$q_{2}$=0.1} \\
\hline
Pb & 105 & 194 & 488 & 4.53$\times10^4$ & 7.2$\times10^3$ & 450 & 0.35 & 0.06 & 0.003 & 0.48 & 0.076 & 0.00477 \\
\hline
Bi & 119 & 220 & 554 & 3.11$\times10^4$ & 4.9$\times10^3$ & 309 & 0.08 & 0.013 & 0.0008 & 0.09 & 0.0143 & 0.00089 \\
\hline
Yb & 120 & 222 & 558 & 3.04$\times10^4$ & 4.8$\times10^3$ & 302 & 0.35 & 0.055 & 0.0035 & 0.38 & 0.06 & 0.0038 \\
\hline
Pt & 240 & 444 & 1116 & 3.8$\times10^3$ & 601 & 38 & 0.72 & 0.11 & 0.007 & 0.96 & 0.15 & 0.01 \\
\hline
Pd & 274 & 506 & 1275 & 2.55$\times10^3$ & 404 & 25 & 0.72 & 0.11 & 0.007 & 0.95 & 0.15 & 0.01 \\
\hline
Y & 280 & 518 & 1302 & 2.39$\times10^3$ & 379 & 24 & 0.17 & 0.027 & 0.002 & 0.17 & 0.027 & 0.002 \\
\hline
Zn & 327 & 604 & 1521 & 1.5$\times10^3$ & 238 & 15 & 1.16 & 0.18 & 0.01 & 1.69 & 0.27 & 0.017 \\
\hline
Mn & 410 & 758 & 1907 & 762 & 121 & 7.6 & 0.08 & 0.013 & 0.0008 & 0.072 & 0.011 & 0.0007 \\
\hline
Ti & 420 & 776 & 1954 & 708 & 112 & 7 & 0.46 & 0.073 & 0.005 & 0.23 & 0.04 & 0.002 \\
\hline
Ni & 450 & 832 & 2093 & 576 & 91 & 5.7 & 0.91 & 0.14 & 0.009 & 1.43 & 0.23 & 0.014 \\
\hline
Fe & 470 & 869 & 2186 & 506 & 80 & 5 & 0.80 & 0.13 & 0.008 & 1.02 & 0.016 & 0.01 \\
\hline
Os & 500 & 924 & 2326 & 420 & 66 & 4.2 & 0.88 & 0.14 & 0.009 & 1.10 & 0.17 & 0.011 \\
\hline
Ru & 600 & 1109 & 2791 & 243 & 38 & 2.4 & 1.17 & 0.18 & 0.01 & 1.35 & 0.22 & 0.013 \\
\hline
Cr & 630 & 1165 & 2931 & 210 & 33 & 2 & 0.94 & 0.15 & 0.009 & 0.78 & 0.12 & 0.0077 \\
\hline
Si & 645 & 1192 & 3000 & 196 & 31 & 1.9 & 1.48 & 0.23 & 0.015 & - & - & - \\
\hline
Be & 1440 & 2662 & 6698 & 18 & 2.8 & 0.17 & 2.00 & 0.32 & 0.02 & 3.08 & 0.49 & 0.03 \\
\hline
C & 2230 & 4122 & 10374 & 4.7 & 0.75 & 0.048 & 1.29 & 0.2 & 0.01 & - & - & - \\
\hline
\end{tabular}
\caption{\footnotesize{Chemical elements and their respective Debye temperatures $^{a}${$(K)$}, Specific heat 
$^{b}${$\left(\frac{J}{mol K}\right)$}, thermal conductivity $^{c}${$\left(\frac{W}{cm\dot K}\right)$},
Electrical conductivity $^{d}${$(ohm\cdot cm)^{-1}\times10^{5}$}, for $T = 300K$ and the $q$-deformed values  for
$q_1\to 1$ and $q_2=0.1$, $q_2=0.5$ and $q_2=1$. 
\cite{kit2}}}
\label{tab2}
\end{table}

\begin{figure}[htb!]
\centerline{
\includegraphics[{angle=90,height=7.0cm,angle=270,width=7.0cm}]{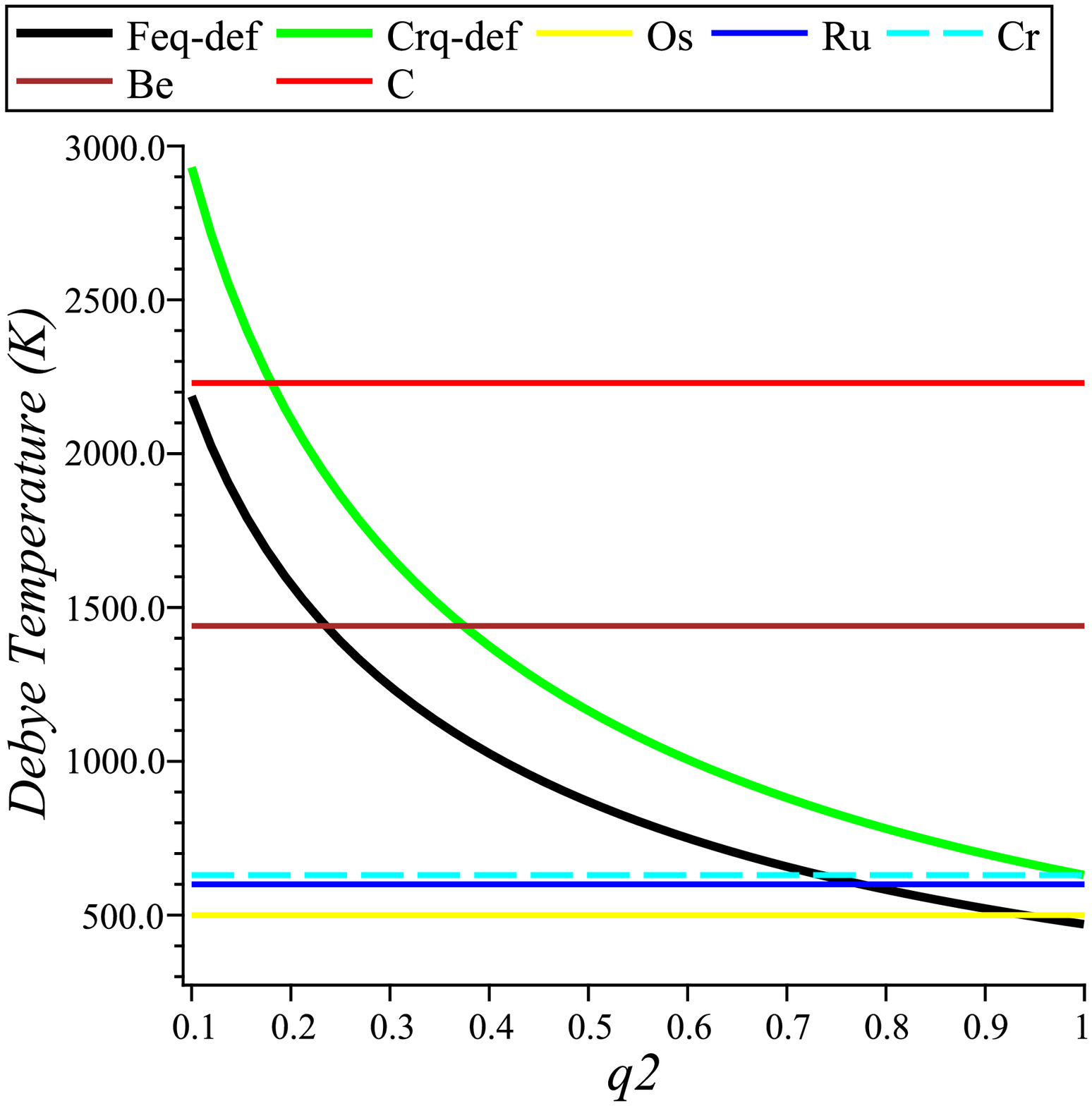}
\includegraphics[{angle=90,height=7.0cm,angle=270,width=7.0cm}]{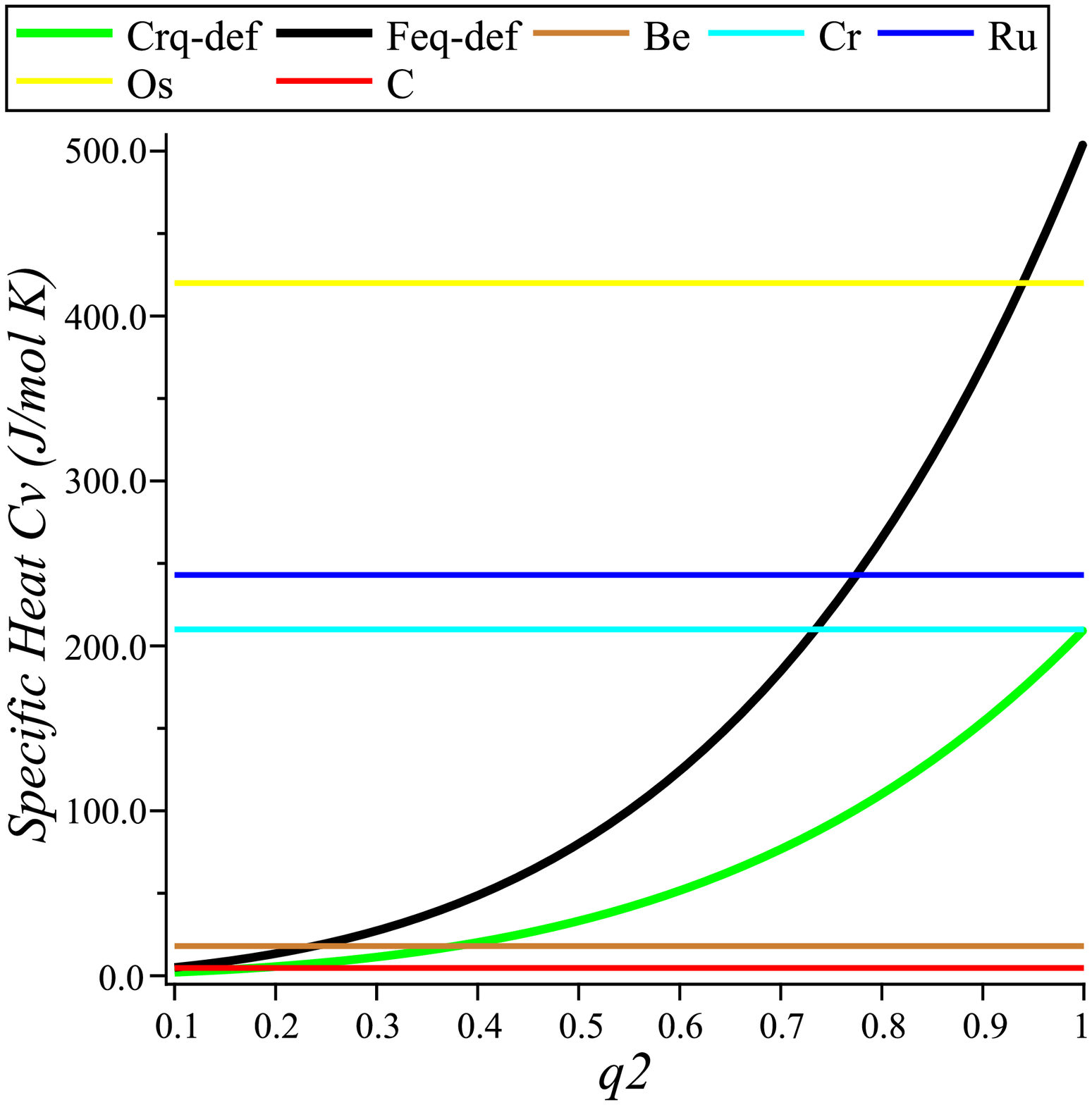}
}\caption{\small{Debye Temperature $\theta_D$ depending on the variation $q_2=0.1,\cdots,1$ and $q_1=1$, 
\textbf{(left)}. Specific Heat $c_{V}$ depending on the variation $q_2=0.1,\cdots,1$ and $q_1=1$, \textbf{(right)}}}
\label{gráficos 16}
\end{figure}

\begin{figure}[htb!]
\centerline{
\includegraphics[{angle=90,height=7.0cm,angle=270,width=7.0cm}]{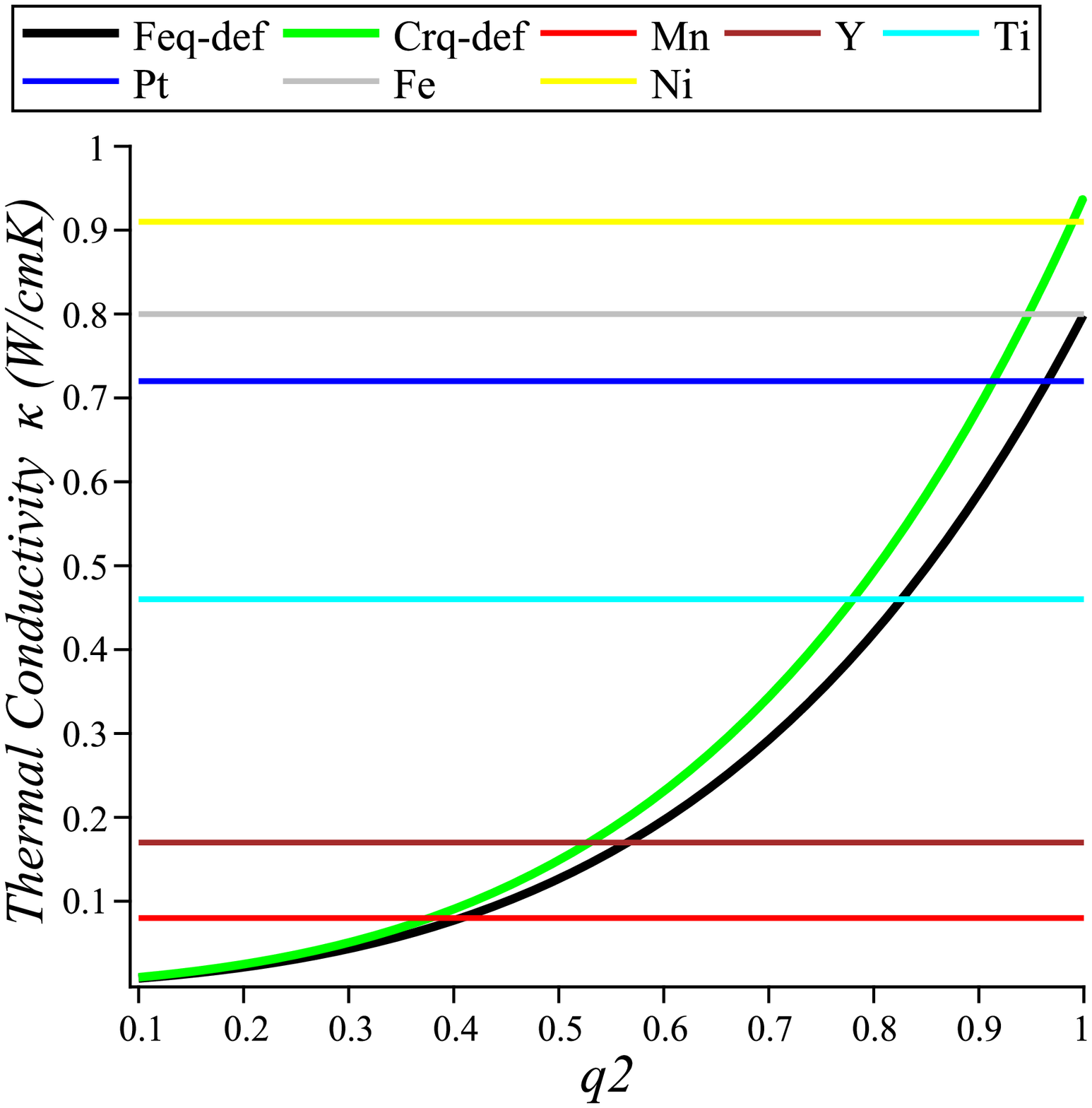}
\includegraphics[{angle=90,height=7.0cm,angle=270,width=7.0cm}]{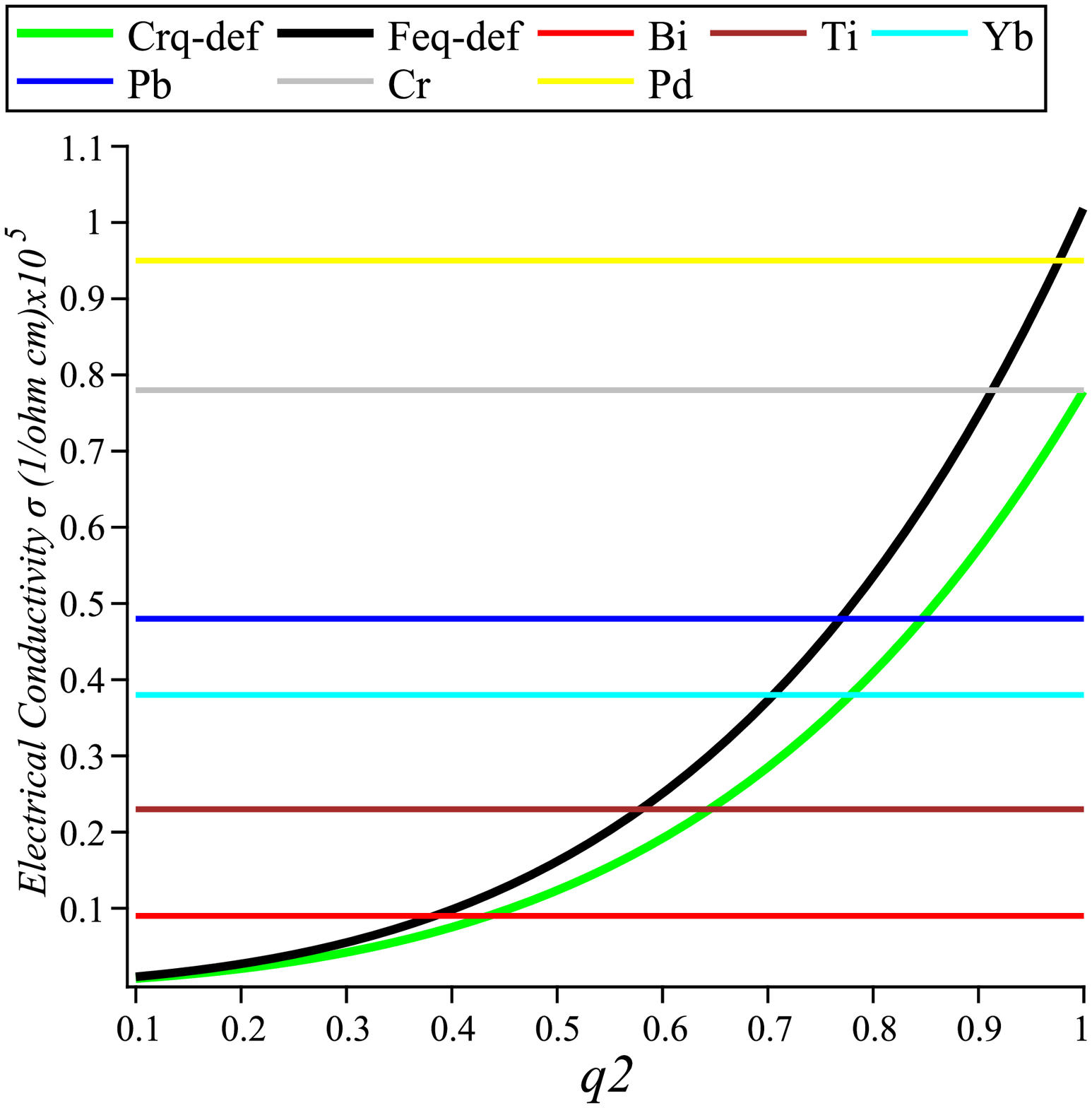}
}\caption{\small{Thermal Conductivity $\kappa$ depending on the variation $q_2=0.1,\cdots,1$ and $q_1=1$, 
\textbf{(left)}. Electrical Conductivity $\sigma$ depending on the variation $q_2=0.1,\cdots,1$ and $q_1=1$, \textbf{(right)}}}
\label{gráficos 18}
\end{figure}

For illustration purposes, we choose iron (Fe) and chromium (Cr), two materials that can be employed in many areas of interest. 
In Figs.(\ref{gráficos 16}, \ref{gráficos 18}) 
we present deformed values of $Fe$ ($Fe_{q-def}$) (black) and $Cr$ ($Cr_{q-def}$) (green), for values 
$q_1=1$ and $q_2=0.1,\cdots,1$, where for this range we assume the maximum deformation ($q_1=1$ and $q_2=0.1$) 
and the pure element (bulk) ($q_1=1$ and $q_2=1$). The other elements are represented by colors and indicated in the very figure.

On the left side of the Fig.(\ref{gráficos 16}), we can observe that before reaching their limits, black and green 
curves can assume the values of Debye temperatures ($\theta_D$) of other elements. The $Fe_{q-def}$ e.g., 
equates to: beryllium (Be) when $q_2\approx 0.23$, chromium (bulk) (Cr) $q_2\approx 0.75$ and osmium (Os) $q_2\approx 0.94$. 
On the right, we have the behavior of the curves obtained for the specific heat $c_V$. We note that the behavior is quite 
different from the previous curves $\theta_D$, i.e., the curves start at lower values (maximum deformation) until they reach 
their pure values. Having $Fe_{q-def}$ as an example again, it is possible to see, 
as it reaches the value of specific heat capacity of all the elements, including $Cr$ (bulk) when $q_2\approx 0.74$.
\begin{figure}[h!]
\centerline{
\includegraphics[{angle=90,height=7.0cm,angle=270,width=7.0cm}]{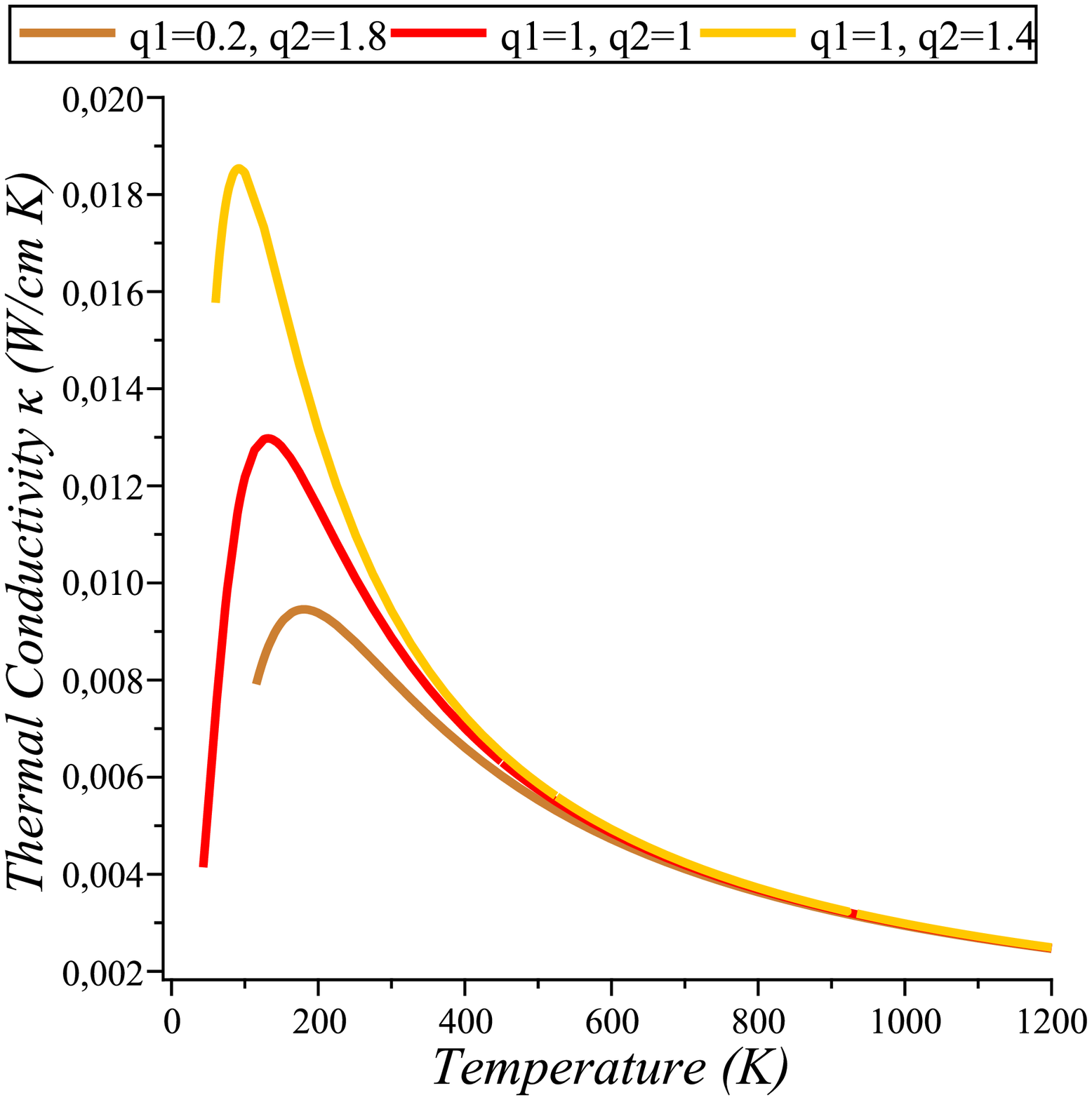}
}\caption{\small{Thermal conductivity $\kappa$ of $Fe$ as a function of temperature in the range of $T=0,\cdots,1200 K$, 
and  values for $q_1$ and $q_2$ given explicitly.}}\label{gráficos 20}
\end{figure}

On the left side of the Fig.(\ref{gráficos 18}), we have the $Cr_{q-def}$ taking over values: 
manganese (Mn) when $q_2\approx 0.4$, titanium (Ti) with $q_2\approx 0.77$ and  $Fe$ bulk $q_2\approx 0.95$. 
On the right, we have the behavior of the curves obtained for the electrical conductivity ($\sigma$), where we observe that the 
ytterbium (Yb) e.g., has its value reached by $Fe_{q-def}$ for $q_2\approx 0.7$ and the $Cr_{q-def}$ $q_2\approx 0.78$. Notice
that in the present case $q_2$ develops an effect of impurity of the material. Such that the more $q$ approaches zero the less thermal and 
electric conductivities approaches zero too. This is in accord with the experimental measures of conductivity of some good conductor that reduces
e.g., its electrical conductivity by doping it with impurities.

Let us now return to Eq.(\ref{eq12.5}), where we have the complete Debye function $D(\alpha_{0_{q_1,q_2}})$ to see the 
behavior of the thermal conductivity. Thus, in the Fig.(\ref{gráficos 20}), we show a comparison to the thermal conductivity 
$\kappa$ for a pure and impure material. 
We have the thermal conductivity as a function of temperature (T) for $Fe$ (bulk) and a combination of the values $q_1$ and 
$q_2$ for the $Fe_{q-def}$ (impure). 
Therefore, we have the $Fe$ (red) when $q_1=q_2=1$ (pure), and when $q_1=0.2$ and $q_2=1.8$, we have a similar behavior to silicon 
(Si) (golden), whose value for the Debye temperature is $\theta_D = 645 K$. Finally, for a combination of values $q_1=1$ and $q_2=1.4$ 
we have a curve that is similar to that of zinc (Zn) (orange), with $\theta_D = 327 K$.

We note that there are a large number of combinations of  $q_1$ and $q_2$ parameter values to be tested. 
In Tab.~(\ref{tab2}) we show only two options ($q_1=1, q_2=0.1$) and ($q_1=1, q_2=0.5$) and Figs.(\ref{gráficos 16}, \ref{gráficos 18}) it is shown that 
one pure material gets impurities by doping, for instance, it may present properties of other \cite{kit2}.
The results of this study with two deforming parameters differ from the results previously obtained in \cite{bri2} by considering only one parameter. Now is
clear there exists another one parameter that can play a different role of the other.

\section{Conclusions}
\label{con}

We apply the Fibonacci oscillators through the energy spectrum in the Einstein solid and thus expanded to the Debye model, 
where our results show that the ($q_1,q_2$)-deformed Debye temperature, specific heat, thermal and electrical conductivity
of `deformed' chemical elements can assume similar values of other pure elements.
The results obtained in our study show that by inserting two deformation parameters $q_1$ and $q_2$, rather than of a parameter $q$, 
increases the adjustment range, i.e., we can have different combinations of values as present in Fig.~(\ref{gráficos 20}).
The existence of more degrees of freedom as in the present case of two deformation parameters, $q_1$ and $q_2$, can be well associated 
with different types of deformations related to two distinct phenomena of disorders or impurities such as, for instance, one due to
pressure generating disorders and other due to doping, respectively.

\section*{Acknowledgments}

We would like to thank CNPq, CAPES, and PNPD/PROCAD-CAPES, for partial financial support.

\end{document}